\documentclass{article}
\usepackage{spconf,amsmath,graphicx,amsfonts}
\usepackage{marvosym}
\usepackage{booktabs}
\usepackage{multirow}
\usepackage[table,xcdraw]{xcolor}

\newcommand{\etal}{\emph{et al. }}

\title{Frequency-Aware Re-parameterization for Over-fitting Based Image Compression}
\name{Yun Ye\textsuperscript{\Letter}, Yanjie Pan, Qually Jiang, Ming Lu, Xiaoran Fang, Beryl Xu}
\address{Intel Corporation, China}

\begin{document}
%
\maketitle
\begin{abstract}
Over-fitting-based image compression requires weights compactness for compression and fast convergence for practical use, posing challenges for deep convolutional neural networks (CNNs) based methods. This paper presents a simple re-parameterization method to train CNNs with reduced weights storage and accelerated convergence. The convolution kernels are re-parameterized as a weighted sum of discrete cosine transform (DCT) kernels enabling direct optimization in the frequency domain. Combined with L1 regularization, the proposed method surpasses vanilla convolutions by achieving a significantly improved rate-distortion with low computational cost. The proposed method is verified with extensive experiments of over-fitting-based image restoration on various datasets, achieving up to -46.12\% BD-rate on top of HEIF with only 200 iterations.

\end{abstract}
\section{Introduction}
\label{sec:intro}

Image compression using deep CNNs has been rapidly developed in recent years. Most existing works focus on learning a representation that can be efficiently compressed for transmission and storage \cite{8693636}. This kind of methods require training on large scale datasets and may not perform well if the image to be compressed is not from the same distribution of training data. To avoid the potential bad performance on out-of-distribution samples as well as to better utilize the fitting ability of deep CNNs, it's a natural idea to over-fit particular data and transmit both the representation and the weights bond to it. Over-fitting based methods have received much less attention compared to learned image compression due to the limitations on both weights storage and training computation cost. Existing over-fitting based methods all focus on reducing the weights storage for efficient transmission and storage. To the best of the authors' knowledge, the second limitation on training efficiency has never been studied. 

\begin{figure}[t]
	\begin{center}
		\includegraphics[width=0.85\linewidth]{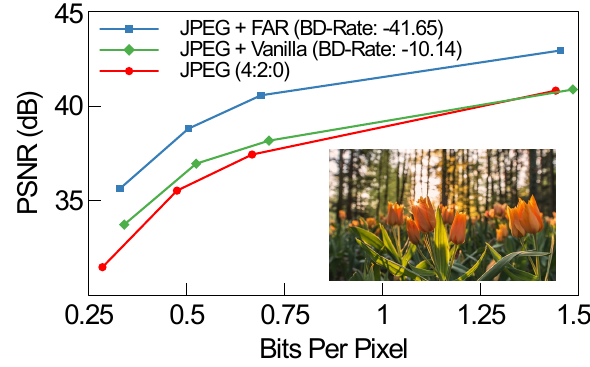}
	\end{center}
	\vspace{-0.5cm}
	\caption{R-D curve of an image from CLIC professional.}
	\label{fig:rdc}
	\vspace{-0.45cm}
\end{figure}

This work aims to tackle both the limitations for the first time, and explores a different direction from prior works. Inspired by the energy compaction property of discrete cosine transform (DCT) \cite{rao1990discrete}, we propose Frequency-Aware Re-parameterization (FAR) that re-parameterizes the convolution kernels by the weighted sum of DCT kernels such that the weights of CNNs can be optimized directly in frequency domain. CNNs based on this simple design converges much faster in training by better capturing the high frequency components. Accompanied with L1 regularization, the method achieves both a high weights sparsity and fast convergence without sacrificing much expressive power. Figure \ref{fig:rdc} is an example illustrating the rate-distortion (R-D) curves of peak signal-to-noise ratio (PSNR) comparing FAR and vanilla convolution. The red curve represents an image compressed by JPEG at different qualities. The blue and green curves are the same images restored by over-fitting using FAR and vanilla convolution, respectively. The corresponding BD-rates of both compared to JPEG are shown in the label. We can clearly see that the images restored by FAR has a much better rate-distortion curve than the images restored by vanilla CNNs, indicating much better expressive powers with the same compressed weights sizes. We further verified the advantages of FAR by extensive experiments conducted on various datasets with popular image codecs. For all the datasets and image codecs, FAR outperforms vanilla convolution measured by the BD-rates in terms of PSNR and multi-scale structural similarity index measures (MS-SSIM). 

The contribution of this paper is two-fold: 1) We propose a simple method, called frequency-aware re-parameterization (FAR), to diminish the two limitations for over-fitting based image/video compression; 2) Extensive experiments conducted on various datasets and codecs to verify the effectiveness of the proposed method. 

\section{Related Works}
\label{sec:related}
Most existing over-fitting based methods study fine-tuning from a pre-trained post-filtering network to restore on top of conventional codecs. Lam \etal proposed to compress weight updates instead of the whole weights \cite{Lam_2019_CVPR_Workshops}. The strategy is later extended by fine-tuning only a part of the weights \cite{9897757,9666144,9675360,10.1145/3394171.3413536}. It can also be combined with super-resolution task \cite{Liu_2021_ICCV,Klopp_2021_ICCV} or a jointly learned representation \cite{rozendaal2021overfitting}. Recent works also explored over-fitting without a pre-trained global model. Video restoration by over-fitting is studied in \cite{He_2020_CVPR_Workshops}. Mikami \etal proposed an approach to over-fit an image using a small-scale auto-encoder and verified on large images \cite{9506367}. This work differs from the above by re-parameterizing convolution to improve both the weights compression and training efficiency, and it is supposed to benefit all the works above.

The effectiveness of DCT for NN weights compression has been studied before. Ko \etal implicitly makes use of DCT to compress NN weights as a JPEG image \cite{7926982}. Ulicny \etal use DCT kernels followed by 1$\times$1 convolutions to replace the conventional convolution layer with reduced redundancy \cite{ulicny2019harmonic}. Their following work applies DCT on reshaped and reordered weight tensors for low-rank pruning and compression \cite{9413944}. DCT for both CNNs compression and running efficiency is studied in \cite{NIPS2016_36366388}. FAR shares the weights compression by DCT with previous works. The key difference is that our method optimizes and regularizes weights directly in frequency domain via DCT kernels. 

This work is benefited from DCT not only in compression but also in training convergence. Traditional deep NNs tend to fit lower frequency components first, which is called spectral bias \cite{pmlr-v97-rahaman19a} or F-principle \cite{2020CCoPh..28.1746X}. Our experiments suggest that by training in frequency domain, the over-fitting is less impacted by the spectral bias and results in a much better convergence.

\section{Method}

\subsection{Frequency-Aware Re-parameterization}
\label{sec:far}

\begin{figure}[t]
	\vspace{-0.5cm}
	\begin{center}
		\includegraphics[width=1\linewidth]{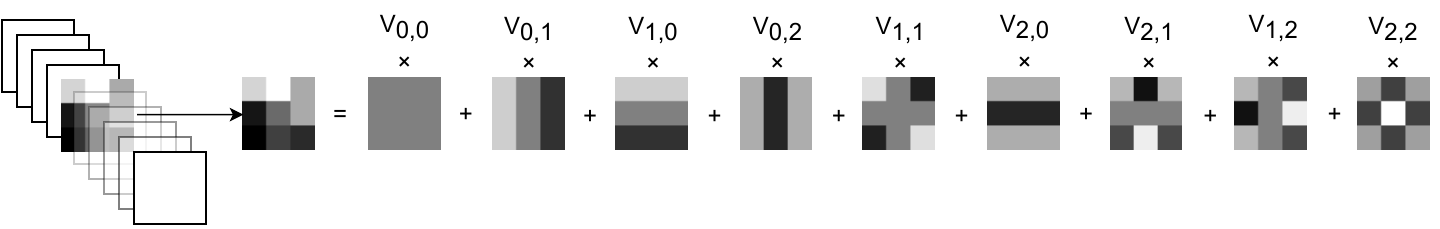}
	\end{center}
	\vspace{-0.5cm}
	\caption{Illustration of re-parameterizing a kernel of a 3$\times$3 convolution as the weighted sum of 9 DCT kernels}
	\label{fig:far}
\end{figure}

The idea is to replace the standard basis of a convolution kernel with basis of frequencies. Denote convolution kernels with $M$ input channels, $N$ output channels, and $H\times W$ size as $\mathcal{K} \in \mathbb{R}^{M \times N \times H \times W}$. A kernel $\mathcal{D}_{ijhw}$ of the orthogonalized DCT-II \cite{rao1990discrete} at subband $i,j \in H \times W$ is expressed as

\begin{equation}\label{expr:kernels}
\hspace{-0.25cm}
\mathcal{D}_{ijhw}=\frac {c_i c_j} {\sqrt {HW}} \cos {\left( \frac {\left( 2h+1 \right)} {2H} i \pi \right)} \cos {\left( \frac {\left( 2w+1 \right)} {2W} j \pi \right)}
\hspace{-0.1cm}
\end{equation}
, where $c_k=1 \text{ if } k=0 \text{ else } \sqrt{2}$. Then the convolution kernels are re-parameterized as the weighted sum over $H \times W$ subbands

\begin{equation}\label{eq:repar}
\mathcal{K}_{mnhw} = \sum_{i=0}^{H-1} \sum_{j=0}^{W-1} \mathcal{V}_{mnij}\mathcal{D}_{ijhw}
\end{equation}, where $\mathcal{V} \in \mathbb{R}^{M \times N \times H \times W}$ is the weight of the convolution using FAR. We illustrate the idea with an example of 3$\times$3 convolution in Fig. \ref{fig:far}. 

\begin{figure}[b]
	\begin{center}
		\includegraphics[width=1\linewidth]{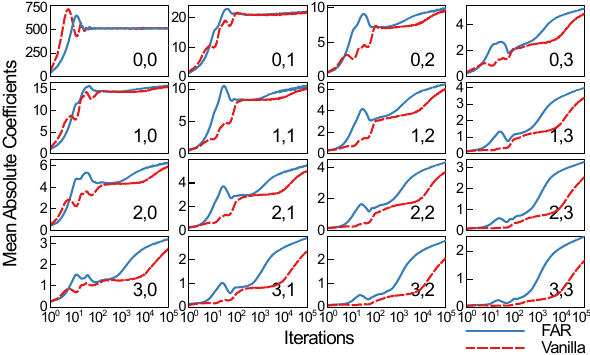}
	\end{center}
	\vspace{-0.5cm}
	\caption{Mean Absolute Coefficients at Different Frequencies.}
	\label{fig:imconvergence}
	\vspace{-0.45cm}
\end{figure}

FAR can be easily implemented with popular deep learning frameworks. It is equivalent to an inverse DCT such that the training is performed in frequency domain. The networks using FAR are trained with L1 regularization, which is verified in training sparse DNNs for compression \cite{NIPS2015_ae0eb3ee}. 

\subsection{Behaviors in Frequency Domain}
\label{sec:freqres}
A toy example comparing the frequency domain behaviors of FAR and vanilla convolution is demonstrated in this section. We train a three-layer CNN with two 512 intermediate channels activated by ReLU to restore an image of 256$\times$256 by over-fitting. The input image is compressed by JPEG at a quality of 15. The network is trained 100,000 iterations using Adam \cite{DBLP:journals/corr/KingmaB14} at a learning rate of 1e-5. The kernel size of both FAR and vanilla convolution is 3$\times$3.

\begin{figure}[t]
	\begin{center}
		\includegraphics[width=1\linewidth]{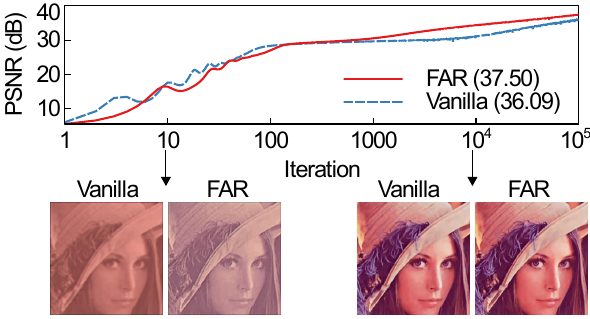}
	\end{center}
	\vspace{-0.5cm}
	\caption{PSNR \& The Restored Images at Different Iterations.}
	\label{fig:visconvergence}
\end{figure}

To check how the different frequency components in the restored image change over training, we decompose the image using a 4$\times$4 DCT and plot the mean absolute values at each subband in Fig. \ref{fig:imconvergence}. The image restored by vanilla convolution exactly follows the spectral bias that the high-frequency components converge much slower than low-frequency components. The FAR based network is less impacted by the spectral bias. The mean absolute coefficients for most high-frequency components converge much faster than vanilla convolution based network. The different behavior can be visually observed in Fig. \ref{fig:visconvergence} that the images restored by FAR are always sharper, particularly at early iterations. 

\begin{figure}[t]
	\begin{center}
		\includegraphics[width=0.9\linewidth]{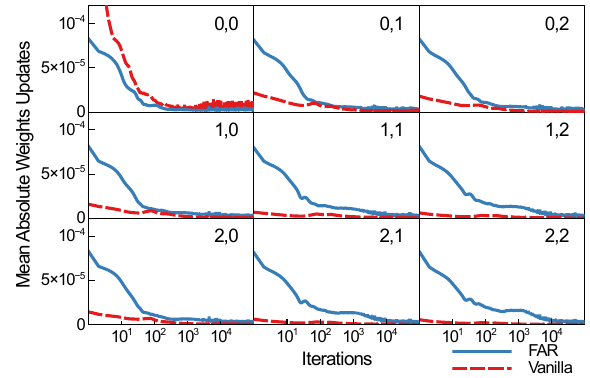}
	\end{center}
	\vspace{-0.5cm}
	\caption{Weights Updates at Different Frequencies}
	\label{fig:weightsupdate}
	\vspace{-0.45cm}
\end{figure}

We also checked the training dynamics by visualizing the weights updates in frequency domain. Figure \ref{fig:weightsupdate} demonstrates the mean absolute change of the DCT coefficients of the 3$\times$3 convolutions at each iteration. We can see that the magnitudes of the weights update of FAR are more uniformly distributed over all subbands, while vanilla convolution updates more to the low-frequency subbands for most iterations. The results demonstrated in Fig. \ref{fig:imconvergence}, Fig. \ref{fig:visconvergence} and Fig. \ref{fig:weightsupdate} all suggest that FAR better learns information distributed accross frequencies. 

\section{Experiments}
\label{sec:exps}

\begin{figure}[t]
	\begin{center}
		\includegraphics[width=1\linewidth]{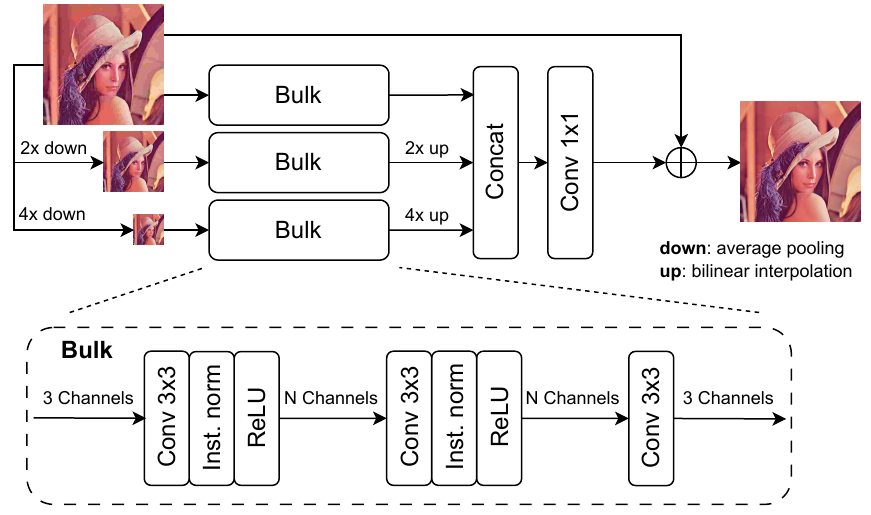}
	\end{center}
	\vspace{-0.6cm}
	\caption{Network architecture for image restoration. The weights of the bulk are shared for inputs at different scales.}
	\label{fig:imrnet}
	\vspace{-0.45cm}
\end{figure}

\subsection{Setup}
\label{sec:setup}
The proposed method is verified with over-fitting based image restoration. Each image compressed by a conventional codec is over-fitted and compared by a network using FAR and vanilla, respectively. Kodak\footnote{https://r0k.us/graphics/kodak}, Tecnick\footnote{https://testimages.org/sampling} and CLIC\footnote {http://compression.cc} 2020 are evaluated with JPEG (cjpeg 9e), HEIF (HEVC, libheif 1.12) and VVC (intra mode, VTM 19.0).

The architecture of the image restoration network is shown in Fig. \ref{fig:imrnet}. We follow \cite{He_2020_CVPR_Workshops} by taking multi-scaled images as inputs. The major difference from both \cite{He_2020_CVPR_Workshops} and the toy example in \ref{sec:freqres} is that the network over-fits the compression residual rather than the raw image. We found that over-fitting residual results in both better compression and convergence for both FAR and vanilla convolution. The bulk of the proposed network is a three-layer CNN with an equal number of channels (N) for the intermediate features. We use instance normalization without affine transform before ReLU as we found it further improves the convergence for both FAR and vanilla convolution. To make a fair comparison, the weight of FAR is initialized by projecting the same weights of vanilla convolution to frequency domain. The default number of channels used for JPEG, HEIF and VVC are 64, 32, and 16, respectively. For Kodak the number is halved as it has smaller images. For JPEG and HEIF the qualities are 15, 40, 65, 90. For VVC the QPs are 37, 32, 27, 22. The pixel formats evaluated are YUV420 and YUV444.

The training objective is the mean-square error between the restored image and the raw image. The network is trained by Adam optimizer for 200 iterations with an L1 penalty of 1e-3 and a linearly decayed learning rate starting from 0.05. After training, the weights are quantized and compressed by DeepCABAC \cite{8970294}. The quantization step size is calculated as $\left | \mathcal{V} \right |_{\max}/L$, where $L$=127 is used in our experiments. Then the quantized weights are loaded back for measuring PSNR, MS-SSIM for the corresponding R-D curves, and BD-rates.

\subsection{Result}
\label{sec:result}

\begin{table*}[t]
	\caption{BD-rates of over-fitting based image restoration. Bold indicates better and red indicates the result is worse than the corresponding codec. CLIC-M and CLIC-P are CLIC mobile and CLIC professional, respectively}
	\vspace{-5mm}
	\label{table:imrestoration}
	\begin{center}
		\setlength\tabcolsep{0.6mm}
\begin{tabular}{r|rr|rr|rr|rr|rr|rr|rr|rr} 
	\hline
	\multicolumn{1}{l|}{\multirow{3}{*}{}} & \multicolumn{8}{c|}{PSNR}                                                                                                                                                                                                         & \multicolumn{8}{c}{MS-SSIM}                                                                                                                                                                                                              \\ 
	\cline{2-17}
	\multicolumn{1}{l|}{}                  & \multicolumn{2}{c|}{Kodak}                             & \multicolumn{2}{c|}{Tecnick}                           & \multicolumn{2}{c|}{CLIC-M}                            & \multicolumn{2}{c|}{CLIC-P}                            & \multicolumn{2}{c|}{Kodak}                                    & \multicolumn{2}{c|}{Tecnick}                           & \multicolumn{2}{c|}{CLIC-M}                            & \multicolumn{2}{c}{CLIC-P}                             \\ 
	\cline{2-17}
	\multicolumn{1}{c|}{}        & \multicolumn{1}{c}{FAR} & \multicolumn{1}{c|}{vanilla} & \multicolumn{1}{c}{FAR} & \multicolumn{1}{c|}{vanilla} & \multicolumn{1}{c}{FAR} & \multicolumn{1}{c|}{vanilla} & \multicolumn{1}{c}{FAR} & \multicolumn{1}{c|}{vanilla} & \multicolumn{1}{c}{FAR} & \multicolumn{1}{c|}{vanilla} & \multicolumn{1}{c}{FAR} & \multicolumn{1}{c|}{vanilla} & \multicolumn{1}{c}{FAR} & \multicolumn{1}{c|}{vanilla} & \multicolumn{1}{c}{FAR} & \multicolumn{1}{c}{vanilla}  \\ 
	\hline
	JPEG 420                               & \textbf{-19.78}         & -6.39                        & \textbf{-41.60}         & -13.50                       & \textbf{-24.22}         & -3.79                        & \textbf{-30.73}         & -11.62                       & \textbf{-14.79}                & -5.07                        & \textbf{-37.52}         & -12.50                       & \textbf{-20.80}         & -4.41                        & \textbf{-26.73}         & -11.44                       \\
	JPEG 444                               & \textbf{-15.56}         & -3.29                        & \textbf{-30.29}         & -2.21                        & \textbf{-19.22}         & -0.72                        & \textbf{-22.83}         & -4.11                        & \textbf{-6.32}                 & \textcolor{red}{0.99}        & \textbf{-26.28}         & -2.37                        & \textbf{-14.53}         & -0.53                        & \textbf{-16.34}         & -3.94                        \\
	HEIF 420                               & \textbf{-19.23}         & -14.63                       & \textbf{-46.12}         & -25.98                       & \textbf{-20.62}         & -10.51                       & \textbf{-30.46}         & -20.25                       & \textbf{-2.01}                 & \textcolor{red}{2.02}        & \textbf{-27.49}         & -4.92                        & \textbf{-7.51}          & \textcolor{red}{3.01}        & \textbf{-20.24}         & -9.13                        \\
	HEIF 444                               & \textbf{-7.94}          & -3.98                        & \textbf{-18.04}         & \textcolor{red}{1.41}        & \textbf{-6.02}          & \textcolor{red}{0.71}        & \textbf{-13.00}         & -3.36                        & \textbf{-0.38}                 & \textcolor{red}{4.42}        & \textbf{-15.39}         & \textcolor{red}{8.05}        & \textbf{-4.38}          & \textcolor{red}{5.88}        & \textbf{-15.84}         & -3.55                        \\
	VVC 420                                & \textbf{-12.30}         & -9.33                        & \textbf{-29.83}         & -20.67                       & \textbf{-16.38}         & -12.11                       & \textbf{-19.95}         & -15.60                       & \textbf{-4.04}                 & -0.85                        & \textbf{-9.80}          & -1.13                        & \textbf{-4.90}          & -0.41                        & \textbf{-10.18}         & -5.77                        \\
	VVC 444                                & \textbf{-0.46}          & \textcolor{red}{1.29}        & \textbf{-4.52}          & \textcolor{red}{2.05}        & \textbf{-2.45}          & -0.01                        & \textbf{-3.20}          & -0.22                        & \textbf{\textcolor{red}{1.43}} & \textcolor{red}{2.86}        & \textbf{-1.20}          & \textcolor{red}{4.58}        & \textbf{-0.88}          & \textcolor{red}{1.56}        & \textbf{-2.39}          & \textcolor{red}{0.39}        \\
	\hline
\end{tabular}
	\end{center}
\end{table*}

The results comparing the BD-rates of FAR and vanilla convolution are listed in Table \ref{table:imrestoration}. FAR overwhelms vanilla convolution in all the evaluations. Vanilla convolution failed to optimize R-D in many cases (red in Table \ref{table:imrestoration}), while FAR only failed to optimize MS-SSIM for VVC(4:4:4) compressed Kodak images. This limits the practical use of small images with highly optimized image codecs. For other cases FAR is promising for further image compression on top of the conventional image codecs. An example R-D curve from the results is shown in Fig. \ref{fig:rdc}.

\begin{figure}[t]
	\begin{center}
		\includegraphics[width=0.95\linewidth]{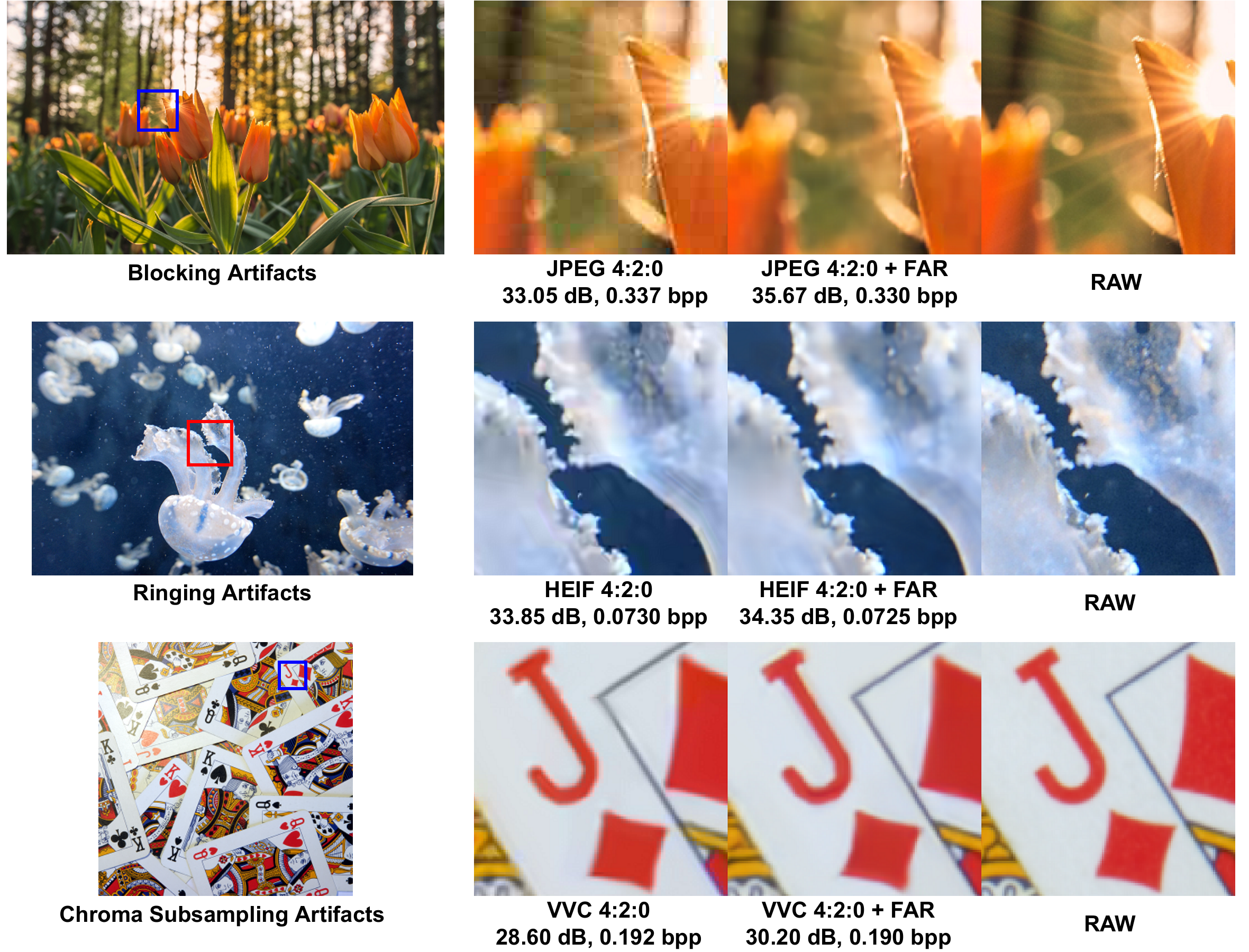}
	\end{center}
	\vspace{-0.5cm}
	\caption{Image samples comparing FAR with image codecs.}
	\label{fig:patchsample}
	\vspace{-0.45cm}
\end{figure}

Figure \ref{fig:patchsample} shows the restored image samples and the same images compressed by the same codec at approximately the same bits per pixel (BPP). We can find that different kinds of artifacts by conventional codecs are removed and it results in much better visual quality. 

\begin{figure}[b]
	\vspace{-0.2cm}
	\begin{center}
		\includegraphics[width=0.85\linewidth]{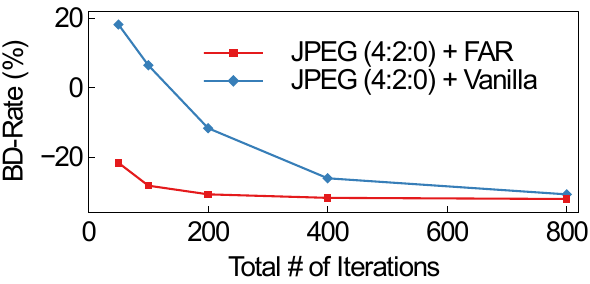}
	\end{center}
	\vspace{-0.5cm}
	\caption{BD-rates of PSNR vs. total number of iterations}
	\label{fig:convergence}
\end{figure}

\subsection{Convergence}

The BD-rates of both FAR and vanilla convolution with different total training iterations on CLIC professional are illustrated in Fig. \ref{fig:convergence}. FAR shows much better convergence over vanilla convolution, particularly at fewer total training iterations. Though the gap between FAR and vanilla convolutions gets smaller as the number of training iterations goes up, FAR is more practical for implementation as it cost much fewer computations in general to achieve the same BD-rate.

\subsection{Ablation Study}
\label{sec:ablation}

L1 regularization is supposed to make both FAR and vanilla convolution weights sparse. To study its impact, we conduct experiments on CLIC professional for both FAR and vanilla convolution. The corresponding BD-rates are summarized in Table \ref{table:ablation}. We found that L1 regularization improves the BD-rates for both as expected. Note that even without L1 regularization FAR outperforms vanilla convolution by a large margin.

\begin{table}[t]
	\caption{BD-rates with \& without L1 regularization.}
	\vspace{+1mm}
	\label{table:ablation}
	\centering
	\begin{tabular}{cc|rr}
		\hline
		L1 & FAR & PSNR   & MS-SSIM  \\ 
		\hline
		$\checkmark$  & $\checkmark$   & \textbf{-30.73} & \textbf{-26.73}   \\
		& $\checkmark$   & -23.38 & -17.34   \\
		$\checkmark$  &     & -11.62 & -11.44   \\
		&     & 4.62   & 3.27    \\
		\hline
	\end{tabular}
    \vspace{-0.5cm}
\end{table}

\section{Conclusion}

A frequency-aware re-parameterization method for convolution has been presented for over-fitting based image compression. It converges fast in training by better capturing high frequency components such that it has the potential for practical use. It achieves much better rate-distortion compared with vanilla convolution based networks, particularly for the case with a few iterations. The method is evaluated with over-fitting based image restoration showing the superiority over vanilla convolution.

\bibliographystyle{IEEEbib}
\bibliography{refs}

\end{document}